\documentclass[5p]{elsarticle}
\usepackage{amsmath}
\usepackage{feynmf}
\usepackage{latexsym}
\usepackage{amssymb}
\usepackage{graphicx}
\usepackage{rotating}
\usepackage{latexsym}
\usepackage{verbatim}
\usepackage{hyperref}
\usepackage{changepage}

 \def\be{\begin{equation}}
 \def\ee{\end{equation}}
 
 \def\a{\alpha}
 
 \def\g{\gamma}

 \def\m{\mu}
 
 \def\t{\tau}
 
 \def\s{\sigma}

 \def\ds#1{#1\kern-1ex\hbox{/}}
 \def\sla{\raise.15ex\hbox{$/$}\kern-.57em}
 \def\nn{\nonumber}
 \def\bea{\begin{eqnarray}}
 \def\eea{\end{eqnarray}}
 \newcommand{\bth}{{\bf 3}}
 \newcommand{\btw}{{\bf 2}}
 \newcommand{\bon}{{\bf 1}}
 \def\QQ{{Q_Q}}
 \def\QU{{Q_{U^c}}}
 \def\QD{{Q_{D^c}}}
 \def\QL{{Q_L}}
 \def\QE{{Q_{E^c}}}
 \def\QHu{{Q_{H_u}}}
 \def\QHd{{Q_{H_d}}}

 \def\({\left(}
 \def\){\right)}
 \def\[{\left[}
 \def\]{\right]}

 \def\ds#1{#1\kern-1ex\hbox{/}}
 \def\sla{\raise.15ex\hbox{$/$}\kern-.57em}

\begin{document}
\begin{frontmatter}
\title{Asymmetry at LHC for an $U(1)'$ anomalous extension of MSSM}

\author[label1]{Francesco Fucito\footnote{Francesco.Fucito@roma2.infn.it}}
\author[label1]{Andrea Mammarella\footnote{Andrea.Mammarella@roma2.infn.it}}
\author[label1]{Daniel Ricci Pacifici\footnote{Daniel.Ricci.Pacifici@roma2.infn.it}}

\address[label1]{Dipartimento di Fisica dell'Universit\`a di Roma ,``Tor Vergata" and
I.N.F.N.~ -~ Sezione di Roma ~ ``Tor Vergata'',\\
 Via della Ricerca  Scientifica, 1 - 00133 ~ Roma,~ ITALY \\
 \vskip 0.5cm
PACS numbers: 12.60.-i, 14.70.Pw \qquad \qquad \qquad~~~~~~~~~~~~~~~~~~~~~~~~~~~~~~~~~~~~~~~~~~~~~~~~~~~~~~~~~~~~ \small{\tt ROM2F/2012/08}}

\begin{abstract}
The measurement of the forward-backward asymmetry at LHC could be an important instrument to pinpoint 
the features of extra neutral gauge particles obtained by an extension of the gauge symmetry group of the standard model.
For definitiveness, in this work we consider an extension of the gauge group of the minimal supersymmetric standard model
by an extra anomalous U(1) gauge symmetry.
We focus on $pp \rightarrow e^+e^-$ at LHC and use four different definitions of the asymmetry obtained implementing
four different cuts on the directions and momenta of the final states of our process of interest. 
The calculations are performed without imposing constraints on the charges of the extra Z's of our model, since the anomaly is cancelled 
by a Green-Schwarz type mechanism. Our final result is a fit of our data with a polynomial in the charges from which to extract the values
of the charges given the experimental result.
\end{abstract}

\end{frontmatter}

\section{Introduction}

One of the most motivated extensions, from a theoretical point of view, of the standard model (SM) and minimal supersymmetric 
standard model (MSSM) of particle physics is obtained by enlarging the gauge group of the theory by admitting extra $U(1)$'s.
Such extensions are natural at low energy for models coming from grand unified theories and string theories (see  \cite{Langacker:2008yv}
for a recent review). In the string inspired scenarios the anomalies of the extra $U(1)$'s are cancelled by the Green-Schwarz mechanism.
To explore such possibility we will use an extension of the MSSM which from now on will be dubbed MiAUMSSM.
An alternative version of this model which admits
spontaneous supersymmetry breaking was also formulated in \cite{Lionetto:2011jp}, but in this work we will use the original formulation of
 \cite{Anastasopoulos:2008jt}.
The phenomenology of the MiAUMSSM has been investigated in different directions.  Assuming that the lightest supersymmetric particle 
(LSP), a candidate for dark matter, comes from the anomalous sector of the model \cite{Fucito:2008ai,Fucito:2011kn}, the relic density of
such LSP was computed and proved to be compatible with  the experimental data of WMAP \cite{Komatsu:2010fb}.
Furthermore in \cite{Lionetto:2009dp} the decays of the next to lightest supersymmetric particle (NLSP) into the LSP has been considered, 
while in \cite{Fucito:2010dj} the features of a possible signature of the model at LHC has been considered by concentrating on a 
particular radiative decay of the NLSP.\\
In this paper we will further develop the phenomenology of the  MiAUMSSM by computing the forward-backward asymmetry which is induced in the final 
states of the process $pp \rightarrow e^+e^-$ by keeping into account the new gauge boson, $Z'$, associated to the extra $U(1)$ gauge symmetry.
The couplings (charges) of this particle to the others present in our model are not fixed by the requirement of gauge anomaly cancellation and can be
determined only by experiment. Our aim is to show that such measurement is feasible and that it can distinguish among the different 
possible scenarios \cite{Langacker:1984dc}.
Since at LHC the colliding beams are made of the same particle, to generate an asymmetry in the final state, some cuts on the parameter
space have to be necessarily performed.  Each possible cut leads to a different definition of the asymmetry. In this work
we will use four different sets of cuts to show that our results are not dependent from these choices.\\
This work is organized as follow: in sec. \ref{sec2} we briefly review the main features of
the model which we are going to study. In sec. \ref{sec3} we will discuss the four different definitions of
the asymmetry we will use: in sec. \ref{sec4} we will describe our calculations and collect the results which are finally
discussed in the conclusions.

\section{Model definition \label{sec2}}
Our model \cite{Anastasopoulos:2008jt} is an extension of the MSSM with an extra $U(1)$. 
The charges of the matter fields with respect to the symmetry groups are given in table 1.
  \begin{table}[h!]
  \centering
  \begin{tabular}[h]{|c||c|c|c|c|}
   \hline & SU(3)$_c$ & SU(2)$_L$  & U(1)$_Y$ & ~U(1)$^{\prime}~$\\
   \hline \hline $Q_i$   & $\bth$       &  $\btw$       &  $1/6$   & $Q_{Q}$ \\
   \hline $U^c_i$   & $\bar \bth$  &  $\bon$       &  $-2/3$  & $Q_{U^c}$
\\
   \hline $D^c_i$   & $\bar \bth$  &  $\bon$       &  $1/3$   & $Q_{D^c}$
\\
   \hline $L_i$   & $\bon$       &  $\btw$       &  $-1/2$  & $Q_{L}$ \\
   \hline $E^c_i$   & $\bon$       &  $\bon$       &  $1$     &
$Q_{E^c}$\\
   \hline $H_u$ & $\bon$       &  $\btw$       &  $1/2$   & $Q_{H_u}$\\
   \hline $H_d$ & $\bon$       &  $\btw$       &  $-1/2$  & $Q_{H_d}$ \\
   \hline
  \end{tabular}
  \caption{Charge assignment.}\label{QTable}
  \end{table}\\
   The gauge invariance of the model implies:
  \bea
   \QU &=& - \QQ - \QHu  \nn\\
   \QD &=& - \QQ + \QHu  \nn\\
   \QE &=& -\QL  + \QHu  \\
   \QHd  &=& - \QHu \nn \label{Qconstraints}
  \eea
Thus, there are only three free charges introduced by the extra symmetry: 
we can choose $\QQ$, $\QL$ and $\QHu$ without loosing generality. The anomalies
induced by this extension are cancelled by the GS mechanism: there are 
no further constraints on the charges.\\
To evaluate the asymmetry associated to the full process $pp\rightarrow e^+e^-$ we have 
performed the calculation of the cross section of the subprocess
$q\bar{q}\rightarrow e^+e^-$, which we report in \ref{crsec}.
In \ref{det} we give details on the convolution of this differential cross section for the specific definitions of asymmetry
we will adopt.
We take the mass of our $Z'$ to be $1.5~\text{TeV}$. There are two main reasons for this choice: on the one hand 
we wanted a sizeble $Z'$ production (see \cite{Anastasopoulos:2008jt}, where there are results for 
a $Z'$ mass of $1~\text{TeV}$).  On the other hand this mass value allows a comparison with the results 
in literature \cite{Zhou:2011dg}.\\
Regardless, our analysis could be repeated for arbitrary value of the  $Z'$ mass.

\section{Asymmetry definition \label{sec3}}

Because the initial $pp$ state is symmetric, the
asymmetry at LHC is zero if we integrate over the whole parameter
space. However the partonic subprocess $q\bar{q}\rightarrow e^+e^-$
is asymmetric. We can keep this asymmetry by imposing  kinematical cuts,
which are anyway inevitable because of the limits imposed by the
detector.
There are many possibilities to perform these cuts and each of them
leads to a different definition of the asymmetry.
In this work we have used the four definitions of the asymmetry, $A_{\rm{RFB}}(Y^{\rm{cut}})$,
$A_{\rm{OFB}}(p^{\rm{cut}}_{z})$, $A_{\rm C}(Y_{\rm C})$ and $A_{\rm E}(Y_{\rm C}) $, 
which are collected  in \cite{Zhou:2011dg}:
\be
A_{\rm{RFB}}=\left. \frac{\sigma(|Y_{
e^-}|>|Y_{{e^+}}|)-\sigma(|Y_{ e^-}|<|Y_{{ e^+}}|)}{\sigma(|Y_{
e^-}|>|Y_{{ e^+}}|)+\sigma(|Y_{e^-}|<|Y_{{
e^+}}|)}\right|_{|Y|>Y^{{cut}}}
\label{arfb} \ee
\be
A_{\rm{OFB}}=\left. \frac{\sigma(|Y_{
e^-}|>|Y_{{ e^+}}|)-\sigma(|Y_{e^-}|<|Y_{{ e^+}}|)}{\sigma(|Y_{
e^-}|>|Y_{{ e^+}}|)+\sigma(|Y_{e^-}|<|Y_{{
e^+}}|)}\right|_{|p_{z}|>p^{\rm{cut}}_{z}}
\label{ao}\ee
\be
A_{\rm C}= \frac{\sigma_{ e^-}(|Y_{e^-}|<Y_{\rm
C})-\sigma_{{ e^+}}(|Y_{{ e^+}}|<Y_{\rm C})} {\sigma_{
e^-}(|Y_{e^-}|<Y_{\rm C})+\sigma_{{ e^+}}(|Y_{{e^+}}|<Y_{\rm C})}
\label{ac} \ee
\be
A_{\rm E}= \frac{\sigma_{e^-}(Y_{\rm C}<|Y_{
e^-}|)-\sigma_{{ e^+}}(Y_{\rm C}<|Y_{{e^+}}|)} {\sigma_{
e^-}(Y_{\rm C}<|Y_{e^-}|)+\sigma_{{e^+}}(Y_{\rm C}<|Y_{{ e^+}}|)}
\label{ae} 
\ee 
where $\sigma$ is the total cross section after integrating with the partonic 
distribution functions (PDFs).\\
The first two asymmetries are defined in the center of mass 
(CM) frame. The forward-backward asymmetry $A_{RFB}$ 
\cite{Langacker:2008yv, Langacker:1984dc},\cite{Petriello:2008zr}\nocite{Cvetic:1995zs,Dittmar:2003ir}-\cite{Godfrey:2008vf} 
has a cut on the rapidity $Y$ of the $e^-/e^+$ pair 
\be
Y=\frac{1}{2}\log \left(\frac{E_{e^-e^+}+p_{z}}{E_{e^-e^+}-p_{z}}\right)
\ee
The one-side asymmetry 
$A_{O}$ \cite{Wang:2010tg,Wang:2010du} has a cut on $p_{z}$, the total momentum associated to the final states ($e^-e^+$) 
moving longitudinally  along the beam direction chosen to be the $z$ axis.\\
In \ref{det} this rapidity will be expressed in the CM in terms of the partonic 
variables $x_{1,2}$. $E_{e^-e^+}$ is  the sum of the energies associated to the two particles. 
The other two asymmetries are defined in the laboratory (Lab) frame.
The variable $Y_{e^{\pm}}$ is the pseudo-rapidity associated to the single particle $e^{\pm}$ and expressed as
\be
Y_{e^{\pm}}=-\log\Big(\tan(\theta^{e^{\pm}}/2)\Big)
\ee
with $\theta^{e^{\pm}}$ the angle of the outgoing fermion with respect to the $z$ axis.
In this case the kinematical cut is over the rapidity in the Lab frame which is denoted by $Y_C$ and which will be introduced in \ref{det}.
The central asymmetry $A_C$ 
\cite{Ferrario:2009ns}\nocite{Kuhn:1998jr,Kuhn:1998kw,Antunano:2007da}-\cite{Ferrario:2008wm} is calculated 
integrating in the angular region centered on the axis orthogonal to the beam, while the edge 
asymmetry $A_E$ \cite{Xiao:2011kp} is defined in the complementary region. \\
For further details, see \ref{det}.

\section{Asymmetry calculation \label{sec4}}
In this work we have calculated the asymmetry in two different ways. First we have used
a numerical code that we have written using Mathematica. This code uses the cross-section 
calculated in 
\ref{crsec} to numerically compute the integrals discussed in \ref{det}.
As a second check we have repeated the same computation using the HERWIG package \cite{Corcella:2000bw,Corcella:2002jc}, that
we have modified to calculate the asymmetry. We have chosen to repeat twice our computation
for two main reasons: the first one is that in this way we can have a cross check between
our results; the second  is that these methods have different peculiarities that we want to 
use. For example, the numerical integration is less computer time consuming for the Mathematica code,
which helps in establishing the dependence of the asymmetry
from the free charges of the model. At the same time the HERWIG package permits to study how the cuts
influence the rate of production of our final state. 
For these reasons we have performed the basic
calculation (i.e. the asymmetry optimization) using both methods.
We remark that all the results that we will show are strongly dependent on the set of PDFs used to
calculate them and that this leads to a systematical error. In the 
following we do not show results for different sets of PDFs. Where the
statistical error is concerned, we have estimated it using the formula \cite{Zhou:2011dg}: 
\be
err \equiv \sqrt{\frac{4 N_F N_B}{N^3}}\simeq \frac{1}{\sqrt{\mathcal{L} \s}} \label{staterr}
\ee where $N_{F/B}$ are the forward/backward events, $N$ is the total number of events and $\mathcal{L}$ is
the luminosity. In the following we show the estimated errors for the asymmetry definitions
keeping $\mathcal{L}=100~fb^{-1}$.\\
We aim to use the asymmetry to distinguish our model from the MSSM or other models which include an extra $U(1)$.
In the following we will perform the asymmetry
calculation around the peak region, that is for
$M_{Z'}-3 \Gamma_{Z'} < M_{e^+e^-}<M_{Z'}+3 \Gamma_{Z'}$, where $\Gamma_{Z'}$ is the total decay rate of the $Z'$. 
As we remarked in \ref{det}, this 
determines the integration domain, that is $(M_{Z'}-3 \Gamma_{Z'})^2 < s <(M_{Z'}+3 \Gamma_{Z'})^2$.
We also compare our results with the ones obtained for the Sequential
Standard Model (SSM), in which there is an extra $Z'$ boson which has the same couplings to fermions such as the $SM$ $Z$ boson \cite{Langacker:2008yv,Altarelli:1989ff}. See section \ref{comp}
for further details on the corresponding Lagrangian and the values of the quantum numbers.

\subsection{Optimized asymmetry}
\begin{figure}[h!]
 \includegraphics[scale=0.65]{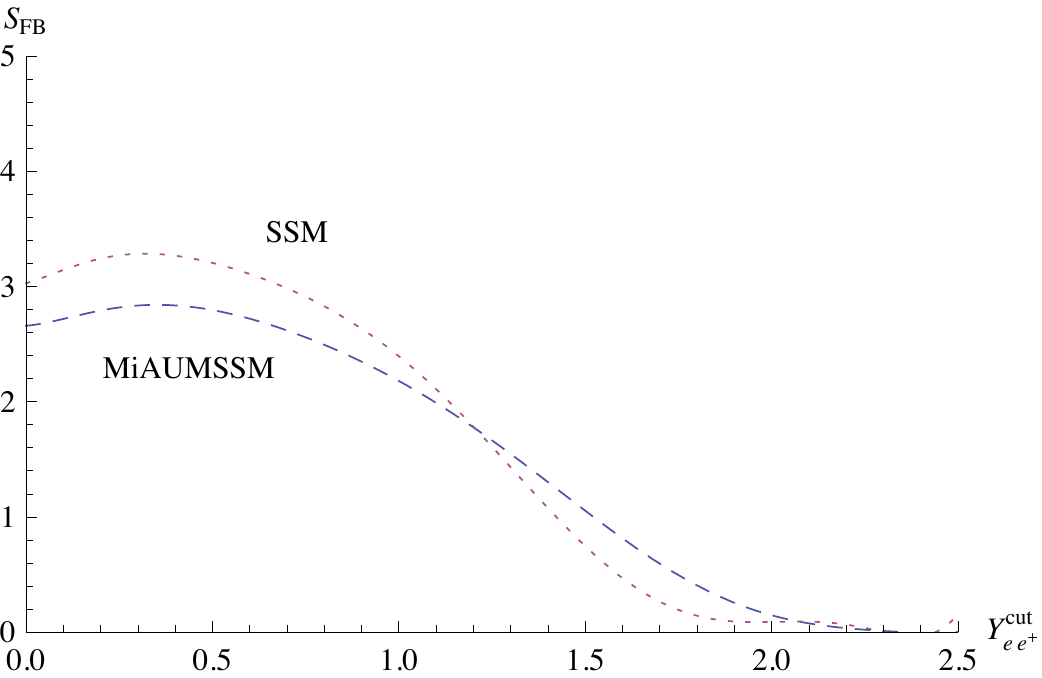}
 \includegraphics[scale=0.65]{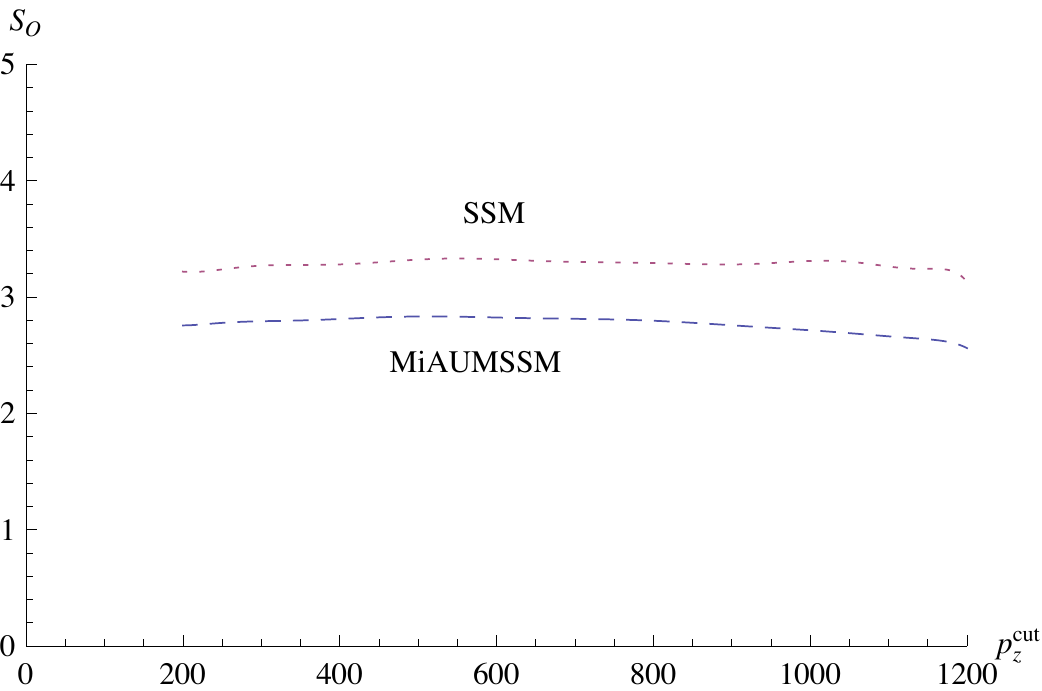}
 \includegraphics[scale=0.65]{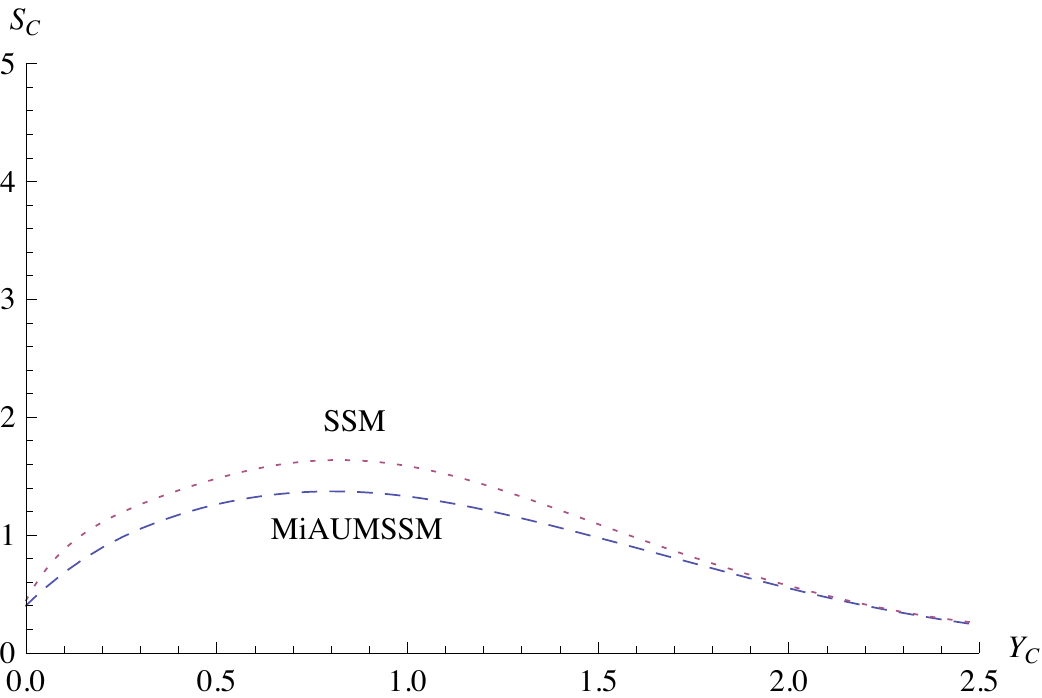}
 \includegraphics[scale=0.65]{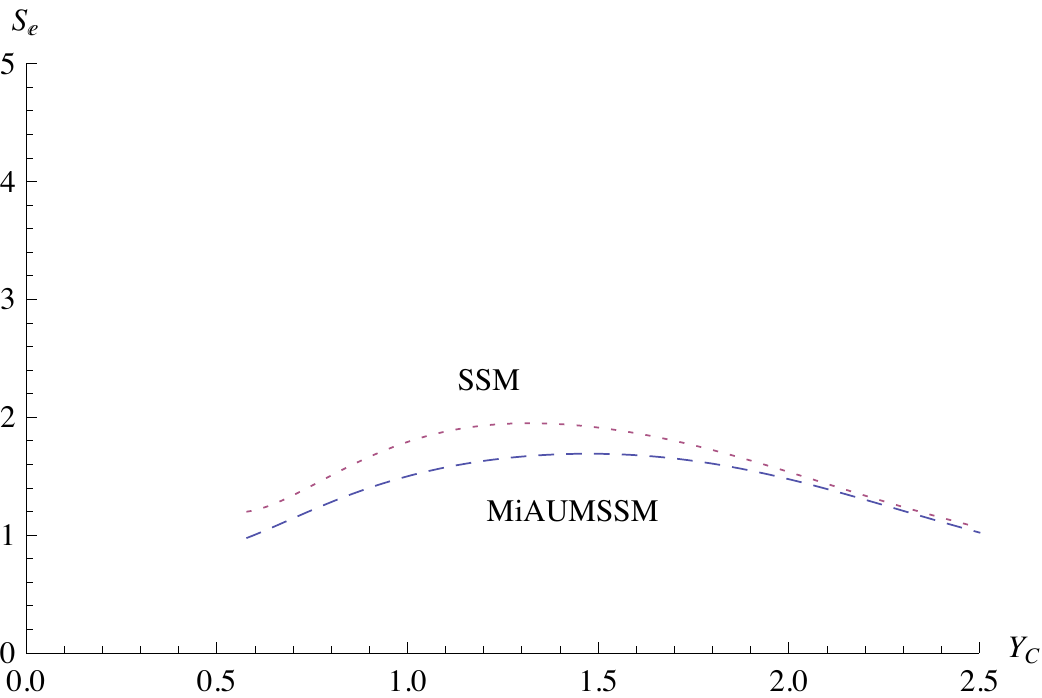}
\caption{Significance as a function of the corresponding cut associated to the four
definitions of asymmetry for on-peak events in both the MiAUMSSM and
the SSM (calculated with the HERWIG package). The charges have been fixed: $Q_{H_u}=0.5$, $Q_Q=0.75$ and $Q_L=1$. The shape of
these functions depends on this choice while the position of the peak does not.}
\label{4sig2}
\end{figure}

As shown in \cite{Zhou:2011dg} the asymmetry magnitude is not a good
function to optimize. A better choice is, instead, the 
statistical significance:

\be Sig \equiv A \sqrt{\mathcal{L}\, \sigma} \ee where $A$ can be any of the 
previously defined asymmetries, $\mathcal{L}$ is the LHC integrated
luminosity, which we take to be $100~fb^{-1}$. \\
We have found a good agreement between the results 
obtained by using the Mathematica code and those obtained with the event simulator HERWIG. 
So we are confident that our results are reliable also when
we will use them to calculate other observables, e.g. the dependence from the charges of
the asymmetries and the significancies.
In figure \ref{4sig2} we show the results for the on-peak significance of the 
MiAUMSSM and SSM for
all the definitions of asymmetry that we use.
The best cuts are those that maximize the significance.
For the SSM we find the same values as in \cite{Zhou:2011dg}. 
We list the best cuts of the MiAUMSSM in table \ref{bestcuts}. 
\begin{table*}[t]
 \centering
\begin{tabular}[h]{|c||c|c|c|c|}
 \hline & $A_{RFB}$ & $A_{O}$ & $A_C$ & $A_E$  \\
 \hline \hline Best cut & $Y_{f\bar{f}}^{cut}=0.4$ & $p_{z,f\bar{f}}^{cut}=580~ GeV$ & $Y_C=0.8$ & $Y_C=1.4$ \\
 \hline 
\end{tabular}
\caption{Best cuts for
the on-peak $e^+e^-$ asymmetries.}
\label{bestcuts}
\end{table*} 
As in \cite{Zhou:2011dg}, we expect that the best cuts are nearly 
independent from the charges and depend only on the $Z'$ mass and the partonic
distribution functions. Moreover they are also essentially independent from the specific model chosen as it is 
confirmed by our analysis. As a further check we have performed simulations with the SSM. We have used the same 
settings of \cite{Zhou:2011dg}, obtaining very similar results 
for all the cuts, confirming the reliability of our numerical codes. We used the 
SSM not only for having a check of the validity of our calculations, but also to have 
results that can be compared with those of the MiAUMSSM.

\subsection{Dependence on the charges \label{dep}}
\begin{figure}
\centering
 \includegraphics[scale=0.5]{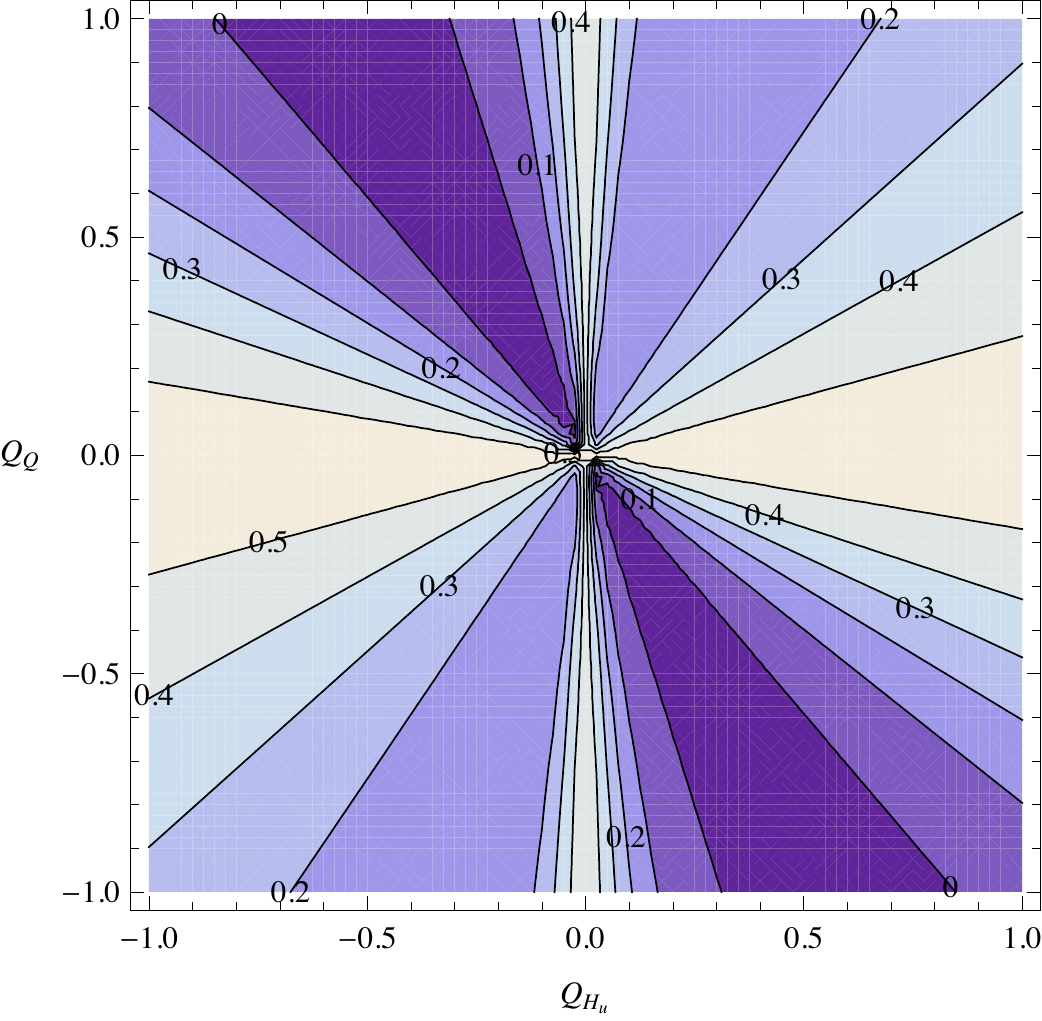}
\caption{Results for the Forward-Backward asymmetry with $Q_{L}=0$ for the best cuts.} 
\label{a1cp}
\end{figure}

\begin{figure}
\centering
  \includegraphics[scale=0.5]{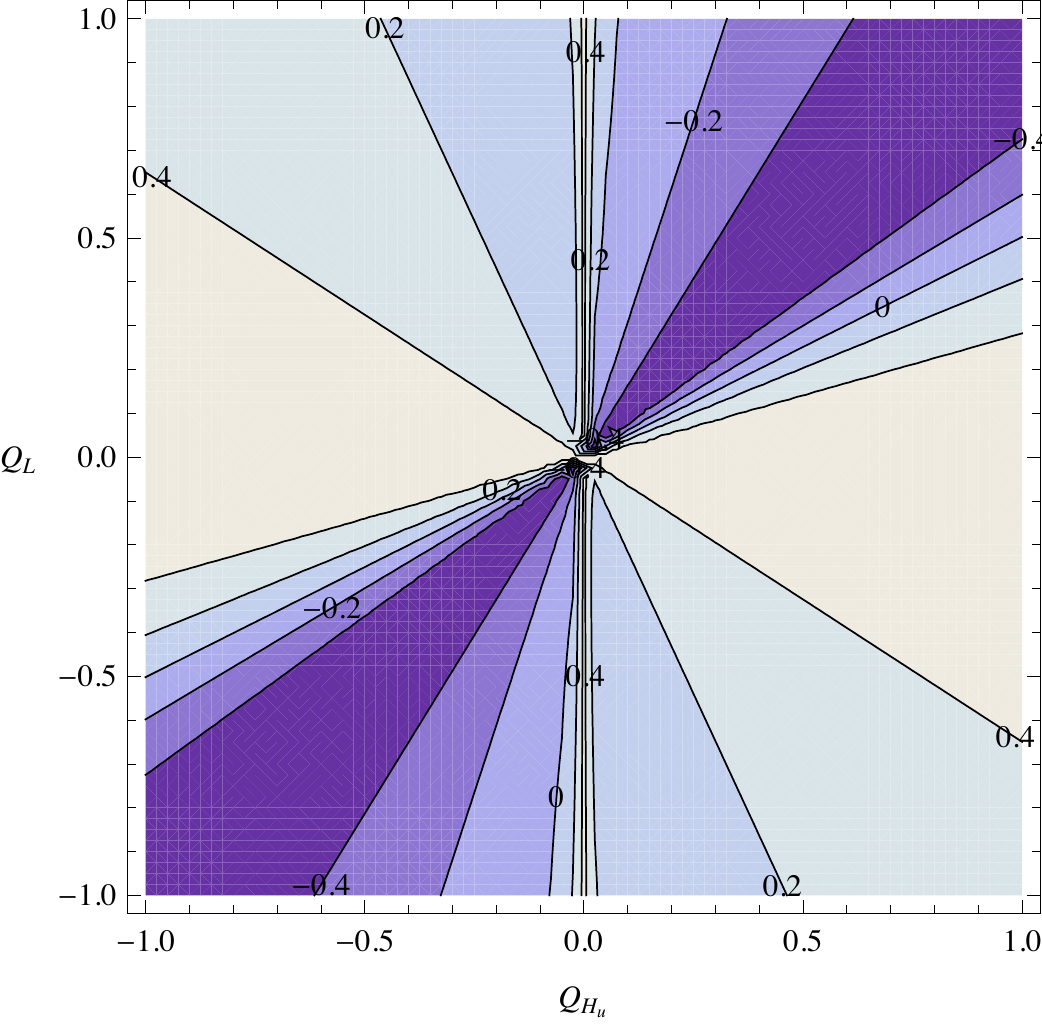}
 \caption{Results for the Forward-Backward asymmetry with $Q_{Q}=0$ for the best cuts.}
 \label{a2cp}
\end{figure}

\begin{figure}
\centering
 \includegraphics[scale=0.5]{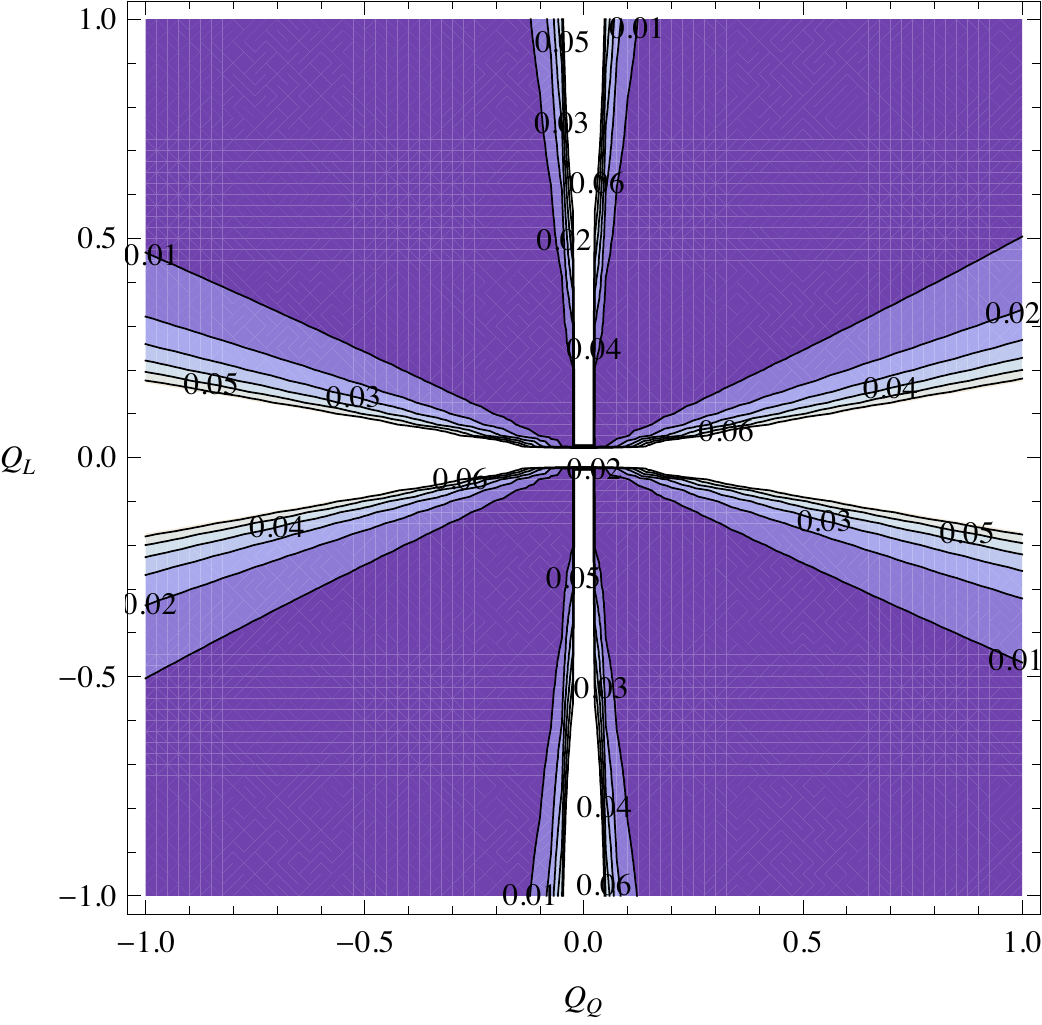}
\caption{Results for the Forward-Backward asymmetry with $Q_{H_u}=0$ for the best cuts. \newline \newline \newline}
\label{a3cp}
\end{figure}

Now we want to use the best cuts previously found to study the asymmetry in function of 
the free charges of our model. We have studied the value of the four asymmetries
keeping alternatively one of the charges fixed to $0$ and varying the others two from 
$-1$ to $1$. We choose these ranges because in the SM all the
charges are of this order. Furthermore in \cite{Fucito:2011kn} we have found that
$-1\lesssim Q_{H_u} \lesssim 1$ for the model to be consistent with the WMAP data on dark matter. 
Then, out of simplicity we have used the same region for the other
two charges. As said, we have obtained the following plots using Mathematica to perform the numerical
integration.
Some of the results we have obtained are showed in figures \ref{a1cp} - \ref{a3cp}. The 
contour plots for the other cases can be found in \cite{Mammarella:2012zd}\\
 In these plots the darkest color areas are those with the lowest absolute values of the asymmetry 
while the greatest values lie in the lightest color region.
In addition, these plots are almost symmetric under exchanges
in the signs of the charges. The contour plots with $Q_{H_u}=0$ 
are almost invariant under the exchange 
$Q_i\to -Q_i$ of the two remaining charges. 
Those with $Q_{L}=0$ or $Q_Q=0$ are almost symmetric only under the 
change of sign of both the two unfixed charges. So the asymmetry as a function of the 
charges must reflect this sort of symmetries in its polynomial dependence on the charges. This 
implies that if we try to fit the asymmetry with a rational function (which is the best choice, given the 
definitions (\ref{arfb}-\ref{ae})) we will have constraints on the coefficients of the fit.

\subsection{Number of events}
We already mentioned that to obtain a non zero asymmetry at LHC we have to impose cuts in the
parameter space. Obviously these cuts will diminish the number of events that we can use to 
measure the asymmetry. It is important to be sure that they do not drastically affect the set of data
we have at our disposal.
To study the ratio between the number of events obtained applying the cuts and the total number of events
expected in our channel of interest ($pp \rightarrow e^+e^-$) we have used the HERWIG package.
We have studied the ratio $N_i/N_{tot_i}$, where $N_i$ is the sum of the forward and backward 
events for the i-th definition of asymmetry and $N_{tot_i}$ is the number of events that we have
generated with HERWIG.
We have performed the calculation of $N_i/N_{tot_i}$ in two cases:
\begin{itemize}
 \item on peak invariant mass, variable cuts
 \item variable invariant mass, fixed cuts
\end{itemize} 
\begin{figure}[h!]
 \includegraphics[scale=0.75]{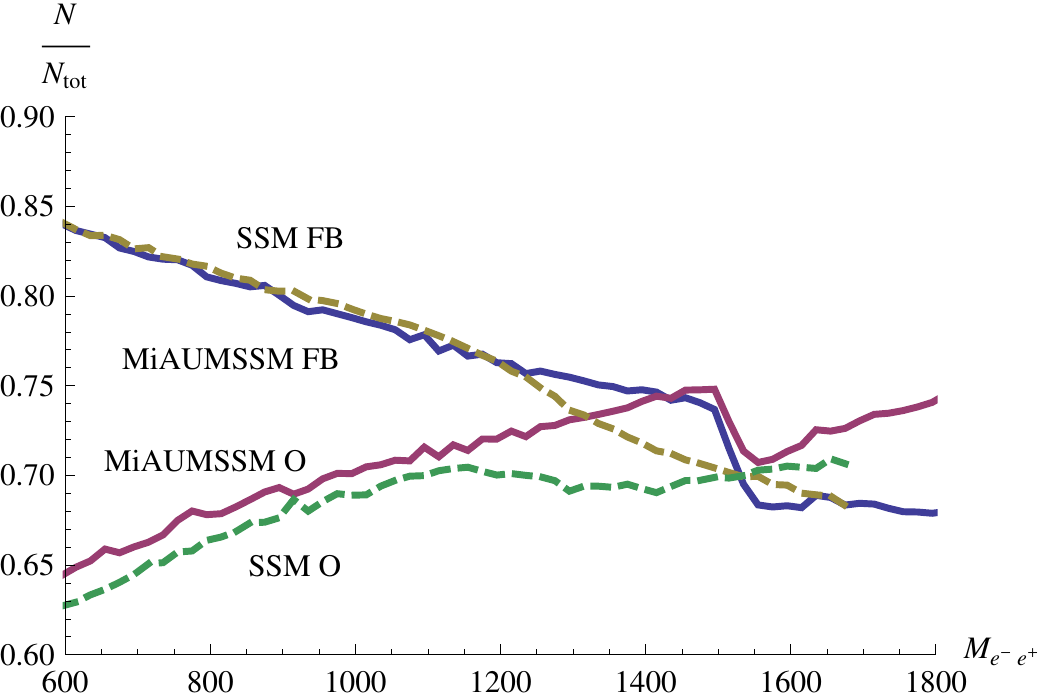}
 \includegraphics[scale=0.75]{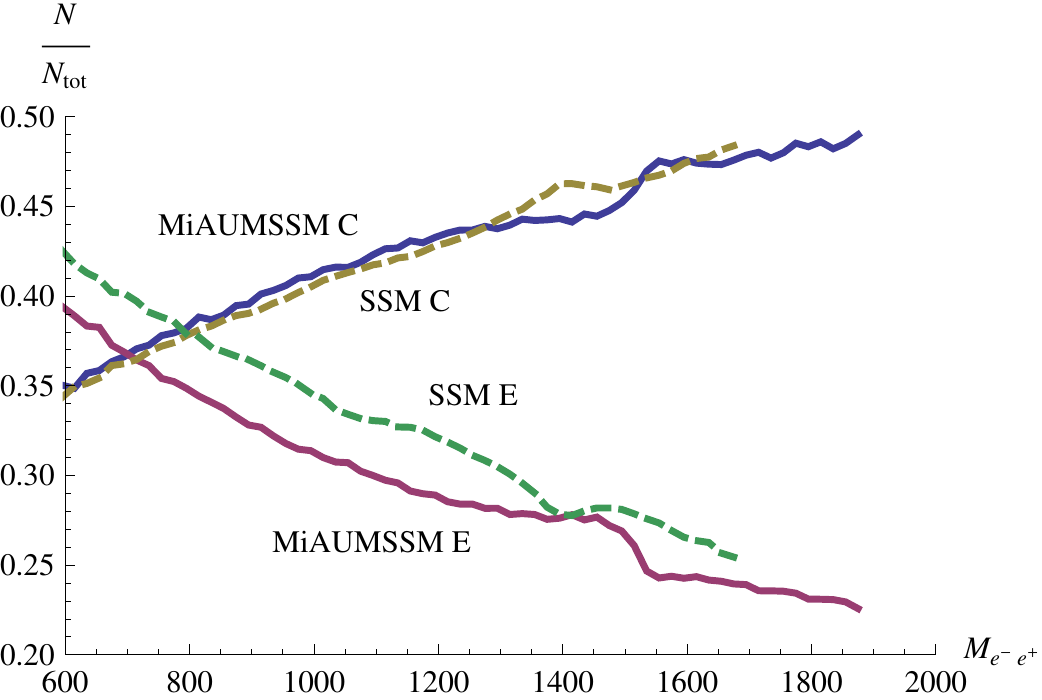}
\caption{$N_i/N_{tot_i}$ as a function of the invariant mass, with the cuts
kept fixed, for the four definitions of asymmetry. In the upper (lower) figure we display the results for the FB and O (C and E)
asymmetries for the MiAUMSSM (thick lines) and the SSM (dashed lines).}
\label{4N1}
\end{figure} Only 
in the case of variable invariant mass with fixed cut it is possible to distinguish the behavior of our model 
from that of the SSM. Therefore we show only the related results in Fig. \ref{4N1}. 
After the implementation of the cuts we are left with the $65-75\%$ of the total number of
events for the FB and O asymmetry, while for the C and E asymmetries we are left with the $40-50\%$ of the events. In both cases these
ratios are good enough to allow the measurement of the observable of interest. 

\subsection{Comparison with other models: LRM and SSM} \label{comp}
\begin{figure}
 \includegraphics[scale=0.45]{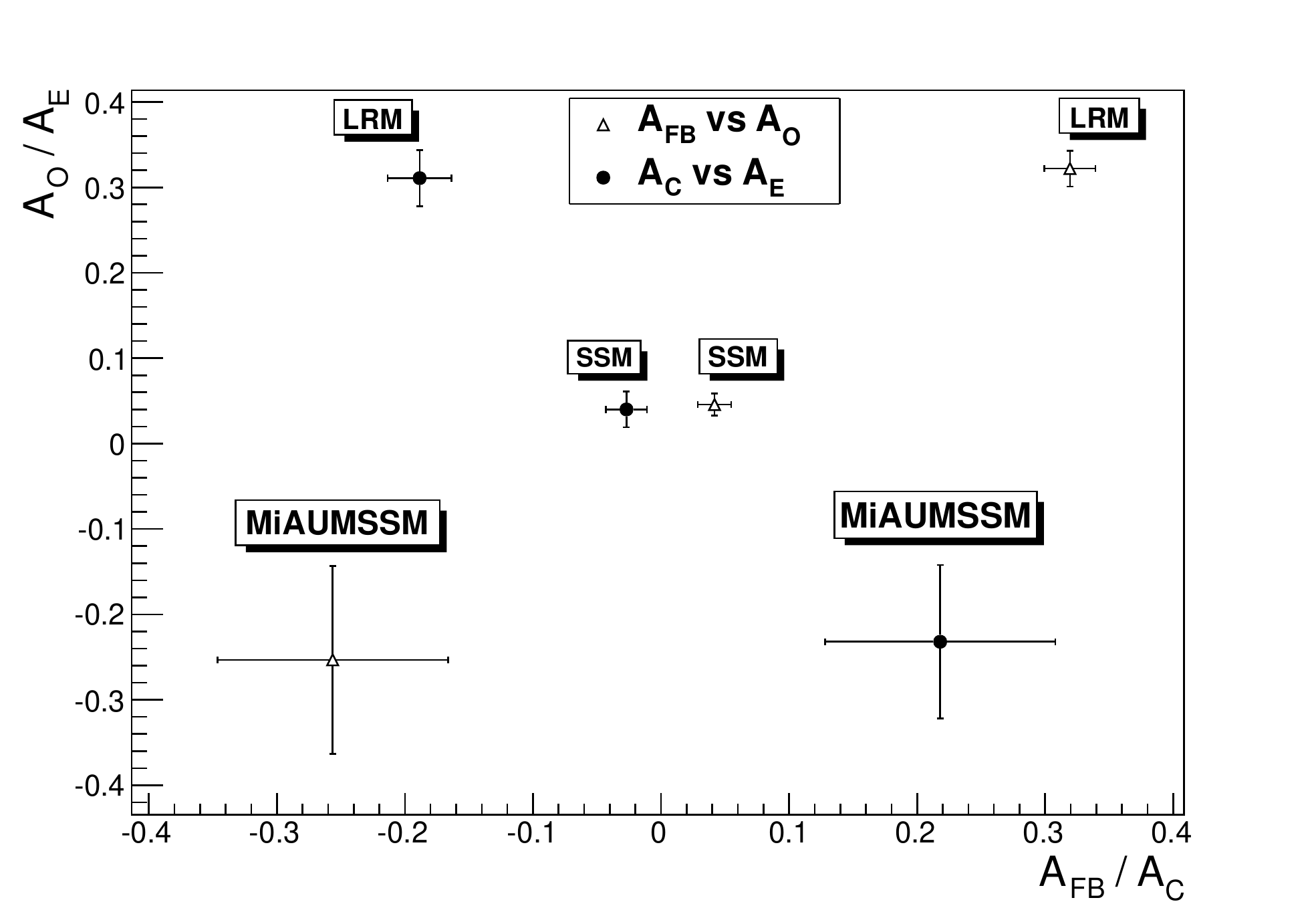}
\caption{The bullets (triangles) define the $A_{FB}$ and $A_O$ ($A_{C}$ and $A_E$) for the three models considered 
in the main text. The charges of the MiAUMSSM are all fixed to $0.5$.}
\label{err1}\end{figure}

\begin{figure*}[t]
 \centering
 \includegraphics[scale=0.55]{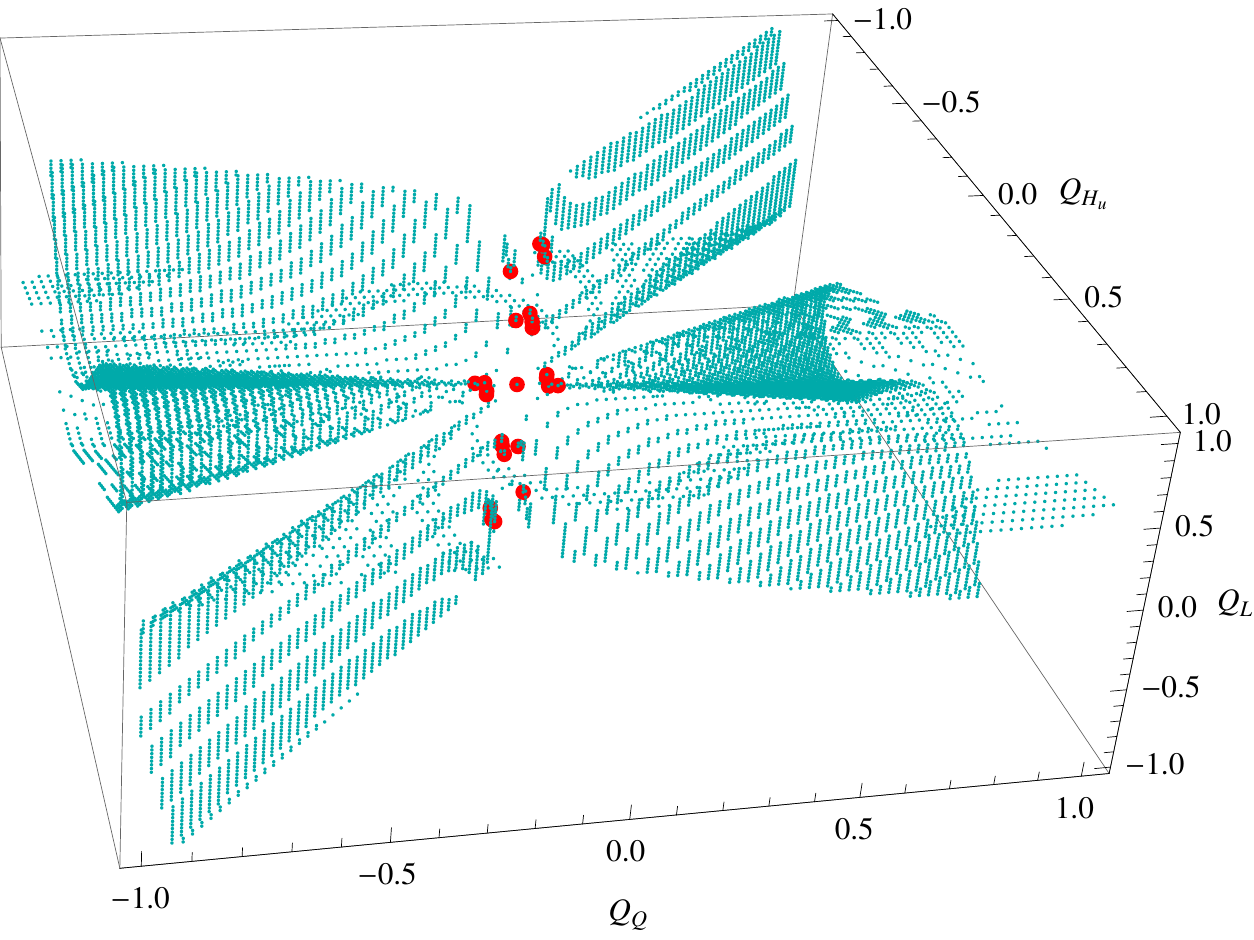} \hspace{1 cm}
  \includegraphics[scale=0.65]{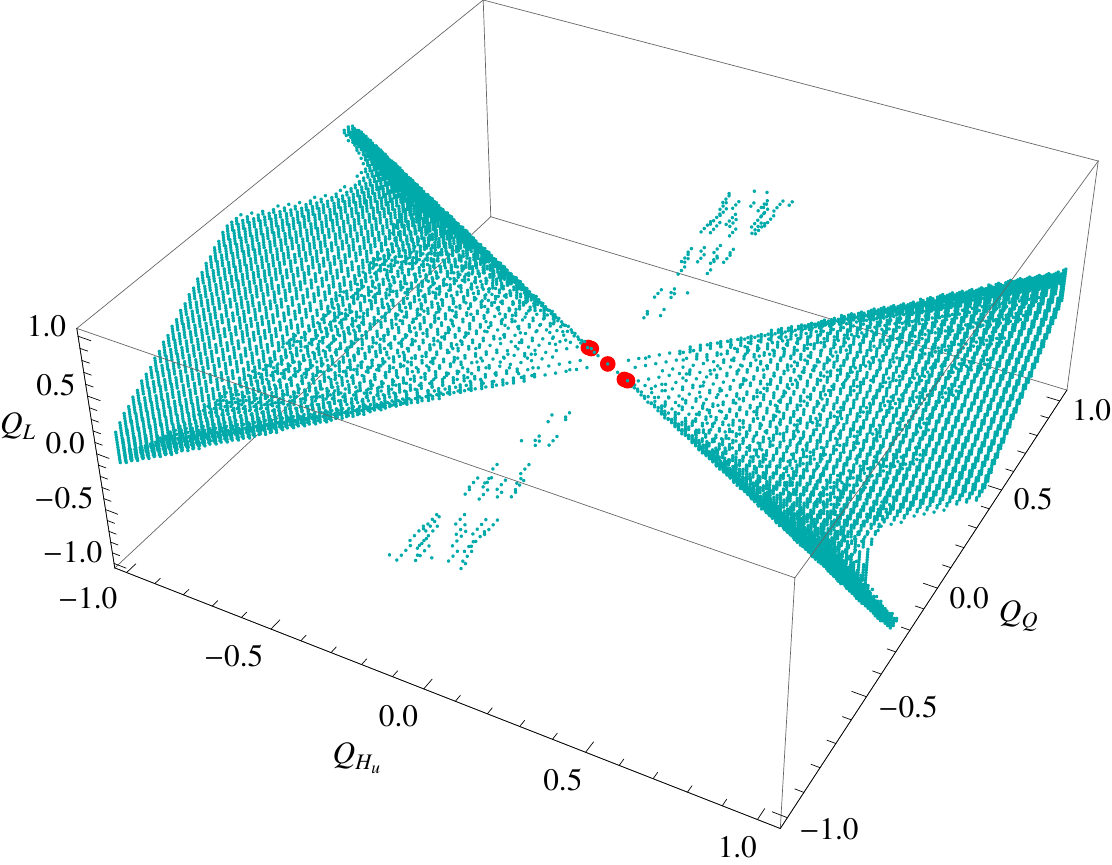}
 \caption{Values of the charges that give a MiAUMSSM asymmetry close to the SSM
asymmetry within errors of 15 (big red dots), 25 $\%$ (left image) and to the LRM within errors of 20 (big red dots), 
25 $\%$ (right image)} \label{SSMim}
\end{figure*}

In this subsection we present a brief analysis of the results for the asymmetry obtained with two well-known
models of extra $U(1)$ extension of the SM: the Left-Right Model (LRM) \cite{Petriello:2008zr,Dittmar:2003ir} 
and the previously mentioned SSM \cite{Altarelli:1989ff}.
We will see that the asymmetry in our anomalous model almost always leads to values
which can be distinguished from those of the LRM and SSM. This implies that a possible future measure
could discriminate among these models.\\
The couplings among the fermions and the $Z'$ for all these models could all be written in the form
\be
g_{Z'} J_{Z'}^{\m} Z'_{\m}=\a \sum_f \bar{\psi_f}\g^{\m}(g_V^f-g_A^f \g_5)\psi_f Z'_{\m}
\ee where:
\be
 \a=\left\{ \begin{array}{cc}
      -\frac{g}{2 \cos \theta_W} & \textrm{SSM} \\
      \frac{e}{2 \cos \theta_W} & \textrm{LRM} \\
      g_0 & \textrm{MiAUMSSM}
     \end{array} \right.
\ee The charges $g_V^f$ and $g_A^f$ are given explicitly in table \ref{couptab}.
$\theta_W$ is the Weinberg angle, defined by $\sin^2 \theta_W=0.231$.
\begin{table*}[t]
\centering
 \begin{tabular}{|c|c||c||c|}
  \hline 
  & SSM & LRM & MiAUMSSM \\
 \hline
 \begin{tabular}{c}
   $f$ \\
   $e,\m,\t$ \\ 
   $u,c,t$ \\
   $d,s,b$ \\ 
  \end{tabular} & 
  \begin{tabular}{c|c}
   $g^f_V$ & $g^f_A$ \\
   $-\frac{1}{2}+2 \sin^2 \theta_W$ & $-\frac{1}{2}$ \\
   $\frac{1}{2}-\frac{4}{3} \sin^2 \theta_W$ & $\frac{1}{2}$ \\  
   $-\frac{1}{2}+\frac{2}{3} \sin^2 \theta_W$ & $-\frac{1}{2}$ \\ 
  \end{tabular} &
\begin{tabular}{c|c}
   $g^f_V$ & $g^f_A$ \\
   $\frac{1}{\a_{LR}}-\frac{\a_{LR}}{2}$ & $\frac{\a_{LR}}{2}$ \\
   $-\frac{1}{3 \a_{LR}}+\frac{\a_{LR}}{2}$ & $-\frac{\a_{LR}}{2}$ \\
   $-\frac{1}{3 \a_{LR}}-\frac{\a_{LR}}{2}$ & $\frac{\a_{LR}}{2}$ \\ 
  \end{tabular} &
\begin{tabular}{c|c}
   $g_V$ & $g_A$ \\
   $Q_L-Q_{H_u}/2$ & $Q_{H_u}/2$ \\
   $Q_Q+Q_{H_u}/2$ & $-Q_{H_u}/2$ \\
   $Q_Q-Q_{H_u}/2$ & $Q_{H_u}/2$ \\ 
  \end{tabular} \\ \hline
 \end{tabular}
\caption{Couplings of the SM fermions to the  $Z'$s for the SSM, LRM and MiAUMSSM models.}
\label{couptab}
\end{table*} 
In the case of the LRM we 
have chosen the so called symmetric version, for which $\a_{LR}=1.59$ \cite{Petriello:2008zr,Dittmar:2003ir}.
Using HERWIG, we have calculated the on-peak asymmetry for these two models. 
Obviously in the case of the MiAUMSSM we do not have a unique value for
the asymmetry, because in the model the charges are not fixed. To show that it is 
possible to distinguish the  MiAUMSSM from the other models we have to estimate the statistical error
in this measurement, by using the formula (\ref{staterr}). 
The exact values depend on the cross section which is model dependent.
Now, if we fix $Q_{H_u}=Q_Q=Q_L=0.5$, the resulting values for the asymmetries associated to
the three models are showed in figure \ref{err1}.
The data plotted in the figure show that it is always possible to discriminate the anomalous model from the non 
anomalous ones.\\
Now we want to stress that the three charges of our model are 
free but the couplings of the fermions to the $Z'$ in the anomalous MiAUMSSM
have a peculiar functional form given in table \ref{couptab}. 
As a consequence it is not possible to match 
the couplings to the extra $Z'$ of the  MiAUMSSM with those of other models.
But, since the four asymmetries have associated 
statistical errors 
we could have a range of values of our three charges where the couplings of the MiAUMSSM 
(and consequently the asymmetries) could be matched with those of the SSM and LRM models within the considered 
errors. In reality this does not happen as we can infer from Fig \ref{SSMim}, where
we consider an error up to $25\%$, much bigger than the expected experimental error. 
Observing the 
amount of points in these figures, it is evident that the SSM is closer to our model than the LRM. This
is the reason why troughout this paper we focus our analysis on the comparison with the SSM.

\subsection{General case}
In this section we want to find the function which describes the asymmetry in terms of the three free charges 
of our model which can assume values between $-1$ and $1$. 
From the cross section of the process, that can be found in \ref{crsec}, 
we can see that the amplitude is proportional to the fourth power of the charges. So, the equations 
(\ref{arfb})-(\ref{ae}) imply that the asymmetry must be a rational 
function in which both the numerator and denominator are fourth grade polynomials in the charges:
\be
 A=\frac{\sum_{i,j,k=0}^n a_{ijk}(Q_{H_u})^i(Q_Q)^j(Q_L)^k}{\sum_{i,j,k=0}^n b_{ijk}(Q_{H_u})^i(Q_Q)^j(Q_L)^k}
\label{AQi}
\ee with $i+j+k= n\le 4$.\\
The apparent symmetries of the contour plots obtained in subsection \ref{dep} imply that the terms of odd 
degree in the charges are suppressed. Then the only relevant terms which do not
contain $Q_{H_u}$  are $Q_Q^2$, $Q_L^2$, $Q_Q^4,~Q_L^4$ and $Q_Q^2 Q_L^2$, while for the terms that do not contain $Q_Q$ or 
$Q_L$ we can also have terms of the form $Q_i^3 Q_j$ or $Q_i Q_j$ where $i,~j$ are the two free charges of each case.
For example, a term proportional to $Q_Q^3 Q_L$ is suppressed, while a term proportional to $Q_Q^3 Q_{H_u}$ 
is present.
Fitting our data for the on peak asymmetries with the functional form (\ref{AQi}), we find the coefficients $a_{ijk}$'s of (\ref{AQi}).  
Then considering only three of the four definitions  we obtain a non-linear system with three equations and three variables 
($Q_{H_u}$, $Q_Q$ and $Q_L$) which could be solved numerically. In this way the asymmetry is useful for 
fixing the values of the $U(1)'$ charges. Moreover, once the values of the three charges are obtained by the previous 
system, the fourth definition of asymmetry can be used as a check for the validity of the model under 
scrutiny. Infact its hypothetical experimental value must be recovered by using (\ref{AQi}) with the values of the 
charges already found, within the considered error (we use the mean relative error (MRE) for each asymmetry definition).
In the \ref{poly} we write a table with the coefficients of the  four fits. As expected, we have found out that the odd 
degree polynomials have negligible coefficients, thus confirming the intuitions stemming from the analysis of the contour plots.
The exactness of the fit is evaluated computing the $R^2$ \footnote{$R^2$ is called coefficient of determination. 
A perfect fit has $R^2=1$.} and the medium relative error for
these results. The results are showed in table \ref{controlfit}, 
attesting the accuracy of the procedure. In particular the $R^2$ value states (as we expected)
that the errors in our fits are almost completely due to the numerical approximations in the calculation
of the integrals and to the uncertainties of the PDFs.
\begin{table}[h]
 \centering
\begin{tabular}[h]{|c||c|c|c|c|}
 \hline & $A_{RFB}$ & $A_{O}$ & $A_C$ & $A_E$  \\
 \hline \hline $R^2$ & $0,999$ & $0,999$ & $0,999$ & $0,999$ \\
 \hline $MRE$ & $0.008$ & $0.009$ & $0.019$ & $0.017$ \\
 \hline
\end{tabular}
\caption{$R^2$ and Medium Relative Error for the polynomial fit of the 
asymmetry with respect to the three charges.}
\label{controlfit}
\end{table} 

\section{Conclusion}
We have numerically calculated the LHC asymmetry of the MiAUMSSM using
four different definitions, namely forward-backward, one-side, central and
edge asymmetries for the process $pp\to e^+e^-$. An analogous study with a complete simulation of the ATLAS detector 
has been carried out for the SM asymmetry \cite{ATLAS:2012as}. The performance of the detector will be unaltered
for a measurement in the TeV range (hoping this to be the scale for the $Z'$) meaning that our results
should be robust even in the case of a complete analysis in which there is a full simulation of the detector. In this respect,
the good agreement, shown in Table 6.4 of \cite{giordano}, for the results coming from Pythia versus those coming from the real measurement
testifies the good degree of precision of the Monte Carlo simulators for these kind of computations. In our case, to achieve 
a statistics comparable to that collected for the case studied in \cite{ATLAS:2012as}, at least 100 fb$^{-1}$ will be needed
and this will be achieved most likely in 2015 when the machine will work at $\sqrt{s}=14-15\, TeV$ after the planned shutdown in 2013/2014.\\
To infer the optimal cuts to use we have maximized the significance related to each asymmetry. We have verified,
as it is expected from \cite{Zhou:2011dg}, that these cuts are nearly independent
from the free charges of our model. Then we have used these optimal cuts to
investigate the asymmetry behavior in function of pairs of free charges,
keeping the third fixed to $0$ to have a graphical representation of
the results. Furthermore we have found that the asymmetry is invariant
under $Q_i \rightarrow -Q_i$.
We have checked that even in presence of the cuts needed to evaluate the asymmetry, the number of left over
events is such to lead to a meaningfull measurement. \\
We have further shown that the MiAUMSSM is distinguishable from the SSM and LRM models, showing examples
of different predictions for the asymmetries which differ at least for a good $20\%$ of their value.
Finally we have studied the four asymmetries as functions of the three
charges and have fitted the results as a rational function of polynomials of degree four in the charges. 
The fit is found to be accurate, with a 
$R^2=0.999$ for all the definitions we have used.

\begin{flushleft}
{\large \bf Acknowledgments}
\end{flushleft}

\noindent The authors would like to thank G. Cattani, G.Corcella,  A. Lionetto, B. Panico, 
A. Racioppi, B.Xiao and Y.-k. Wang
for useful discussions and correspondence during the completion of this paper.

\begin{appendix}

\section{Cross section \label{crsec}}

We have calculated the cross section in the CM for the process $q\bar{q}\rightarrow e^+e^-$ for a general
Drell-Yan interaction in which we can have the  product of diagrams where the 
$\g$, $Z_0$ and $Z'$ can be exchanged . Thus we have six possible terms: $\g\g$, $\g Z_0$, $\g Z'$, 
$Z_0 Z_0$, $Z_0 Z'$ and $Z' Z'$. \\The total amplitude is:
\begin{equation}
|M|^2(q)=\frac{1}{3} \frac{1}{4} \sum_{a,b=\gamma,Z,Z'} {g}_{a}^{2} \, {g}_{b}^{2} \, {M}_{ab} \label{M2sum}
\end{equation} where the fractions $\frac{1}{4}$ and $\frac{1}{3}$ come out from the averages over 
spin and color,  $g_0$ is the coupling associated to the $Z'$. We fix its value
to $0.1$. $M_{ab}$ is the amplitude of each process divided by the couplings:

 \be
  {M}_{ab}=\frac{ 64  \, N_{ab}  \, \Big[(s-m_a^2) \,(s-m_b^2)+
(\Gamma_a \,  m_a \,  \Gamma_b  \, m_b)\Big]}{\Big[(s-m_a^2)^2+(\Gamma_a \, m_a)^2\Big ]\Big[(s-m_b^2)^2+
(\Gamma_b \, m_b)^2\Big]} \nn \ee
\be \Big \{  2 \,  m_e^2 \,  m_q^2 \mathcal{C}^{AAVV}_{e,-} \mathcal{C}^{AAVV}_{q,-}- m_e^2 \, \big (p_{q}\cdot p_{\bar{q}}\big ) \mathcal{C}^{AAVV}_{e,-} \mathcal{C}^{AAVV}_{q,+}\nn\ee
\be+\big[ (p_q \cdot p_{\bar{e}} ) \, (p_{\bar{q}}\cdot p_e )-(p_q\cdot p_e ) \,( p_{\bar{q}}\cdot p_{\bar{e}}) \big] \mathcal{C}^{VAAV}_{e,+} \mathcal{C}^{VAAV}_{q,+}+\nn\ee
\be+\big[ (p_q \cdot p_{\bar{e}} ) \, (p_{\bar{q}}\cdot p_e )+(p_q\cdot p_e ) \,( p_{\bar{q}}\cdot p_{\bar{e}}) \big] \mathcal{C}^{AAVV}_{e,+} \mathcal{C}^{AAVV}_{q,+}+\nn\ee
\be-m_q^2\, (p_e\cdot p_{\bar{e}}) \mathcal{C}^{AAVV}_{e,+} \mathcal{C}^{AAVV}_{q,-} \Big \} 
\label{sigmac}  
\ee defining $ \mathcal{C}^{MNPQ}_{i,\pm}=C_{i,a}^{M}C_{i,b}^{N}\pm C_{i,a}^{P}C_{i,b}^{Q}$, with $i=e,q$ and $M,N,P,Q=V,A$.\\              
In this expression $N_{ab}$ is a multiplicity factor that is equal to $\frac{1}{2}$ if 
the exchanged vector bosons  are identical 
and is equal to $1$ if they are different. The C's are
simply the vector and axial quantum numbers related to the vector bosons: for the $\g$ and the $Z_0$ they are
the usual SM quantum numbers that can be found in \cite{HalzMart}, while the vector and axial
couplings related to the $Z'$ have been calculated in \cite{Mammarella:2012zd} and are showed in table \ref{anochar}.\\
We remark that in this cross section there are only terms of degree four in the powers of the charges.
However, from the equation (\ref{M2sum}) we know that there are different contributions to the total squared amplitude 
of our process. These terms are divided in three types: the term $Z'Z'$, the terms $\g Z'$ and $Z_0 Z'$ and those $\g \g$, $\g Z_0$ and $Z_0 Z_0$ which give contribution of degree four, two and zero in the anomalous charge respectively.
Since we are studying the on-peak region, we expect the contribution from the $Z'Z'$ channel to be dominant
with respect to the others: this is evident from the table of the coefficients showed in \ref{poly}.
Observing the previous formula we can see that all the combinations of $C$'s contain two $C_q$ and two $C_e$
because our elementary process involves two leptons and two quarks. Observing that $Q_Q$ and $Q_L$ are related to 
$C_q$ and $C_e$ respectively, this implies that we cannot have terms of degree larger than two in $Q_Q$ and $Q_L$.
This is verified by our fit, where the coefficients related to this terms are suppressed (see \ref{poly}).
\begin{table}
 \centering
\begin{tabular}[h!]{|c||c|c|}
 \hline & $C^V_{f,Z'}$ & $C^A_{f,Z'}$ \\
 \hline \hline $f=u,c,t$ & $Q_Q+Q_{H_u}/2$ & $-Q_{H_u}/2$ \\
 \hline $f=d,s,b$ & $Q_Q-Q_{H_u}/2$ & $Q_{H_u}/2$ \\
 \hline $f=e,\m,\t$ & $Q_L-Q_{H_u}/2$ & $Q_{H_u}/2$ \\
 \hline 
\end{tabular}
\caption{Vector and axial quantum numbers of the SM fermions with respect to the $Z'$.}
\label{anochar}
\end{table} The differential cross section can be found multiplying for the usual
kinematic prefactor and summing this result over the contribution of the six possible initial quarks:

\begin{equation}
\frac{\partial^2\sigma}{\partial s\partial \cos\theta}\Big|_{CM}=\sum_q \frac{p_e}{32 \, \pi \,  s \, p_q}|M|^2(q)
\end{equation}

\section{Details on the Asymmetry definitions} \label{det}
The explicit expression of eq. (\ref{arfb}) in the CM frame is
 \be
 A_{RFB}=\frac{\int_{C_{cut}}dx_1dx_2\sum_q f_q(x_1)f_{\bar{q}}(x_2)(F-B)}{\int_{C_{cut}}dx_1dx_2\sum_q f_q(x_1)f_{\bar{q}}(x_2)(F+B)}
\label{arfb1} \ee where
 \be
 F=\int_0^1 d\cos\theta \frac{d\sigma(\cos\theta, s)}{d\cos\theta\, ds}~~~~~~~ 
B=\int_{-1}^0 d\cos\theta \frac{d\sigma(\cos\theta, s)}{d\cos\theta\, ds}~~~~~~~
\label{FandB}
 \ee are the forward and backward contributions, respectively. \\ The $f_{q/\bar{q}}(x_i)$ are the
PDFs of $q/\bar{q}$. $C_{cut}$ is the domain
of integration, that depends on the type of asymmetry that we want to calculate according
to the definitions (\ref{arfb}-\ref{ae}).\\ The couples
of variables $(x_1,x_2)$ and $(s,Y)$ are not independent. In fact, their definition is $s=S\, x_1 x_2$ and $Y=\frac{1}{2}\log(x_1/{x_2})$, where $S=(14~TeV)^2$ is the total squared
energy of the accelerator.\\
Since we have calculated the
cross section of the process (see \ref{crsec}) that we are going to study with respect to $s$, we perform
the change of variables $(x_1,~x_2) \rightarrow (s,Y)$ in the expression (\ref{arfb1}).
The Jacobian of this transformation is $J=1/S$. 
 Now we focus on the integral $\int_{-\infty}^{-Y_{cut}} dY$: if we perform the change of variable $Y\to -Y$, because 
of the $Y$ definition, this corresponds to the exchange $x_1\leftrightarrow x_2$ and consequently to 
the exchange of forward with backward ($F\leftrightarrow B$). 
Summarizing
 \bea
 &&\int_{-\infty}^{-Y_{cut}} dY \sum_qf_q(x_1)f_{\bar{q}}(x_2)(F\pm B)=\nn \\
&&\int_{Y_{cut}}^{\infty} dY \sum_qf_q(x_2)f_{\bar{q}}(x_1)(\pm F+ B)
 \eea 
Using this result we obtain the formula that we implemented in the Mathematica evaluation, 
that is:

 \be
  A_{RFB}=\frac{\int ds J \int_{Y_{cut}}^{+\infty}dY \sum_q f^-_{q\bar{q}}(x_1(s,Y),x_2(s,Y))(F-B)} {\int ds J  \int_{Y_{cut}}^{+\infty}dY \sum_q f^+_{q\bar{q}}(x_1(s,Y),x_2(s,Y))(F+B)}   
   \ee
   with
   \be
   f^{\pm}_{q\bar{q}}(x_1,x_2)=f_q(x_1)f_{\bar{q}}(x_2)\pm f_{\bar{q}}(x_1)f_q(x_2)
   \ee 
The formula for $A_O$ is very similar and it is obtained by replacing $Y$ with
$p_z$. This leads to a different cut and a different Jacobian.\\
The expressions for $A_C$ and $A_E$ are also very similar between them and we present them together.
In these two cases the cut is 
performed on the angle $\theta^{cut}$ and therefore on the limits of integration for  F and B. 
These asymmetries are defined in the Lab frame, because the Lorentz transformation from the CM frame 
\textquotedblleft squezees \textquotedblright the final particles \cite{Ferrario:2009ns}. 
 The angles of the outgoing $e^{\mp}$ with respect to the $z$ axis, denoted by $\pm\theta$ respectively in the  CM frame,
 are replaced by $\theta^{e^-}$ and $\theta^{e^+}$ in the Lab frame where the outgoing $e^\pm$ no longer have 
 the same direction.  Therefore in the Lab frame, instead of
the definitions (\ref{FandB}), we have:
 \bea
  \label{FandBlab}
&& ( F/B)_{C}=\int_{-cut}^{+cut}\frac{d\sigma}{d\cos\theta^{e^{(-/+)}} ds}d\cos\theta^{e^{(-/+)}}\nn \\
 && ( F/B)_{E}=\int_{-1}^{-cut}\frac{d\sigma}{d\cos\theta^{e^{(-/+)}} ds}d\cos\theta^{e^{(-/+)}}+\nn \\
 &&\int_{cut}^{+1}\frac{d\sigma}{d\cos\theta^{e^{(-/+)}} ds}d\cos\theta^{e^{(-/+)}}
  \eea
  where the limit of integration $cut$ is defined in terms of $\theta^{cut}$ as $cut=\cos\theta^{cut}$.
Then (\ref{ac}) and (\ref{ae}) become:
  \be
 A_{C/E}=\int ds\int_{s/S}^1dx_1 J \sum_q f_{q}(x_1)f_{\bar{q}}(x_2)\Big(F_{C/E}-B_{C/E}\Big)  
 \ee where $x_2=\frac{s}{S x_1}$.
 $J$ is now the Jacobian of the transformation $(x_1,x_2)\rightarrow (x_1,s)$.\\  
The cut parameter which enters in the definitions (\ref{ac}) and (\ref{ae}) is not directly 
$\theta^{cut}$ but the associated pseudorapidity $Y_C=-\log\Big(\tan(\theta^{cut}/2)\Big)$. 
For further details on the calculations sketched in this Appendix, see \cite{Mammarella:2012zd}.

\section{Coefficients of the polynomial fit \label{poly}}
We have performed a numerical calculation of the asymmetries letting the three charges
vary in the $-1<Q_i<1$ range. Then we have fitted the results with the rational function
(\ref{AQi}). We have found that only the even grade terms contribute to the results, 
so we neglect the odd grade terms.\\
Another point to note is that the formula (\ref{AQi}) implies that all the coefficients $a_{ijk}$ 
and $b_{ijk}$ are defined up to a global multiplicative factor. To permit the comparison among the 
different types of asymmetries we have fixed $a_{400}=1$ (or $a_{400}=-1$ for the C asymmetry that
assumes opposite sign with respect to the others). However, if such type of models will be
discovered at the LHC, this value will be fixed differently to match the experimental results.
The coefficients values for our choice are listed in table
\ref{tab:fits1}. Note that this table contains only the statistical error and not the systematic error
due to the choice of the PDFs.
   \begin{table*}[h]
       \centering
\begin{tabular}{|c||c|c|c|c|c|}
\hline
               &          $A_{RFB}$              &           $A_O$               &           $A_C$               &         $A_E$\\
\hline \hline
    $a_{000}$  & $(-0.52\pm0.02) \times 10^{-6}$ & $(-0.31\pm0.04)\times10^{-6}$ & $(1.18\pm0.04)\times10^{-6}$  & $(0.86\pm0.18)\times10^{-6}$   \\
\hline
    $a_{200}$  & $(82\pm3) \times 10^{-6} $      & $(58\pm5)\times10^{-6}$       & $(-17\pm5)\times10^{-6}$      & $(140\pm21)\times10^{-6}$ \\
\hline
    $a_{020}$  & $(9.9\pm1.5)\times 10^{-6}$     & $(9\pm2)\times10^{-6}$        & $(-27\pm3)\times10^{-6}$      & $(12\pm11)\times10^{-6}$ \\
\hline
    $a_{002} $ & $(5.2\pm1.4)\times 10^{-6}$     & $(2\pm2)\times10^{-6}$        & $(11\pm2)\times10^{-6}$       & $(61\pm10)\times10^{-6}$ \\
\hline
    $a_{110}$  & $(5\pm3)\times 10^{-6}$         & $(4\pm5)\times10^{-6}$        & $(19\pm6)\times10^{-6}$       & $(113\pm24)\times10^{-6}$ \\
\hline
    $a_{101} $ & $(-18\pm4)\times 10^{-6}$       & $(-6\pm6)\times10^{-6}$       & $(-148\pm7)\times10^{-6}$     & $(-269\pm27)\times10^{-6}$  \\
\hline
    $a_{011} $ & $(-80\pm3)\times 10^{-6}$       & $(-65\pm5)\times10^{-6}$      & $(34\pm5)\times10^{-6}$       & $(-177\pm22)\times10^{-6}$\\
\hline
    $a_{400}$  & $1 (\text{fixed})$              & $1 (\text{fixed})$            & $-1 (\text{fixed})$           & $1 (\text{fixed})$ \\
\hline
    $a_{040}$  & $0.015638\pm0.000004$           & $0.015634\pm0.000006$         & $-0.015444\pm0.000007$        & $0.01552\pm0.00003$ \\
\hline
    $a_{004}$  & $0.000969\pm0.000002$           & $0.000967\pm0.000004$         & $-0.000984\pm0.000004$        & $0.000964\pm0.000018$ \\
\hline
    $a_{310} $ & $0.88011\pm0.00005$             & $0.87460\pm0.00007$           & $-0.85003\pm0.00008$          & $0.8573\pm0.0003$ \\
 \hline
    $a_{220}$  & $0.01525\pm0.00004$             & $0.01544\pm0.00006$           & $-0.01565\pm0.00007$          & $0.0163\pm0003$ \\
\hline
    $a_{130} $ & $-0.000729\pm0.000019$          & $-0.00061\pm0.00003$          & $0.00077\pm0.00003$           & $-0.00113\pm0.00013$ \\
\hline
    $a_{301}$  & $-1.99340\pm0.00005$            & $-1.99305\pm0.00008$          & $1.99360\pm0.00009$           & $-1.9930\pm0.0004$  \\
\hline
    $a_{211}$  & $-1.75973\pm0.00010$            & $-1.74845\pm0.00016$          & $1.69973\pm0.00018$           & $-1.7141\pm0.0007$  \\
\hline
    $a_{121}$  & $-0.00388\pm0.00007$            & $-0.00403\pm0.00011$          & $0.00429\pm0.00012$           & $-0.0052\pm0.0005$\\
\hline 
    $a_{031}$  & $0.00165\pm0.00002$             & $0.00128\pm0.00003$           & $-0.00188\pm0.00004$          & $0.00244\pm0.00015$  \\
 \hline  
    $a_{202}$  & $0.00456\pm0.00010$             & $0.00438\pm0.00016$           & $-0.00463\pm0.00018$          & $0.0048\pm0.0007$ \\
\hline
    $a_{112}$  & $-0.00049\pm0.00009$            & $-0.00044\pm0.00014$          & $0.00013\pm0.00015$           & $-0.0011\pm0.0006$ \\
\hline
    $a_{022}$  & $0.00773\pm0.00003$             & $0.07740\pm0.00005$           & $-0.00778\pm0.00006$          & $0.0068\pm0.0002$ \\
\hline  
    $a_{103}$  & $-0.001244\pm0.000013$          & $-0.00120\pm0.00002$          & $0.00136\pm0.00002$           & $-0.00170\pm0.00009$ \\
\hline   
    $a_{013}$  & $0.000181\pm0.000015$           & $0.00023\pm0.00002$           & $-0.00014\pm0.00003$          & $0.00047\pm0.00011$ \\
\hline \hline
    $b_{000}$  & $(-1.19\pm0.06)\times10^{-6}$   & $(-0.70\pm0.09)\times10^{-6}$ & $(-3.19\pm0.12)\times10^{-6}$ & $(2.2\pm0.4)\times10^{-6}$   \\
\hline
    $b_{200}$  & $(121\pm7)\times10^{-6}$        & $(181\pm12)\times10^{-6}$     & $(-88\pm16)\times10^{-6}$     & $(-637\pm57)\times10^{-6}$ \\
\hline
    $b_{020}$  & $(23\pm4)\times10^{-6}$         & $(21\pm6)\times10^{-6}$       & $(74\pm7)\times10^{-6}$       & $(29\pm28)\times10^{-6}$ \\
\hline
    $b_{002} $ & $(12\pm3)\times10^{-6}$         & $(5\pm5)\times10^{-6}$        & $(-29\pm6)\times10^{-6}$      & $(154\pm25)\times10^{-6}$ \\
\hline
    $b_{110}$  & $(-58\pm26)\times10^{-6}$       & $(51\pm41)\times10^{-6}$      & $(-282\pm52)\times10^{-6}$    & $(2392\pm204)\times10^{-6}$ \\
\hline
    $b_{101} $ & $(-116\pm13)\times10^{-6}$      & $(-207\pm20)\times10^{-6}$    & $(408\pm27)\times10^{-6}$     & $(-392\pm100)\times10^{-6}$  \\
\hline
    $b_{011} $ & $(85\pm39)\times10^{-6}$        & $(-82\pm62)\times10^{-6}$     & $(-104\pm79)\times10^{-6}$    & $(-1120\pm306)\times10^{-6}$\\
\hline
    $b_{400}$  & $1.90571\pm0.00004$             & $1.90585\pm0.00006$           & $2.28541\pm0.00008$           & $2.1465\pm0.0003$ \\
\hline
    $b_{040}$  & $0.036021\pm0.000010$           & $0.036021\pm0.000016$         & $0.04190\pm0.00002$           & $0.03945\pm0.00008$ \\
\hline
    $b_{004}$  & $0.002232\pm0.000006$           & $0.002229\pm0.000009$         & $0.002670\pm0.000012$         & $0.00245\pm0.00005$ \\
\hline
    $b_{310} $ & $1.40007\pm0.00018$             & $1.3899\pm0.0003$             & $1.3347\pm0.0004$             & $1.2650\pm0.0015$ \\
 \hline
    $b_{220}$  & $3.8269\pm0.0003$               & $3.8323\pm0.0004$             & $4.5879\pm0.0006$             & $4.338\pm0.002$ \\
\hline
    $b_{130} $ & $-0.00386\pm0.00018$            & $-0.0033\pm0.0003$            & $-0.0035\pm0.0004$            & $-0.0103\pm0.0014$ \\
\hline
    $b_{301} $ & $-3.79461\pm0.00014$            & $-3.7943\pm0.0002$            & $-4.5504\pm0.0003$            & $-4.2761\pm0.0011$  \\
\hline
    $b_{211}$  & $-2.7984\pm0.0004$              & $-2.7775\pm0.0007$            & $-2.6661\pm0.0009$            & $-2.533\pm0.003$  \\
\hline
    $b_{121}$  & $-7.5882\pm0.0007$              & $-7.5998\pm0.0011$            & $-9.1030\pm0.0014$            & $-8.592\pm0.005$\\
\hline 
    $b_{031}$  & $0.0081\pm0.0002$               & $0.0061\pm0.0003$             & $0.0098\pm0.0005$             & $0.0168\pm0.0017$  \\
 \hline 
    $b_{202}$  & $3.8004\pm0.0003$               & $3.8021\pm0.0004$             & $4.5567\pm0.0006$             & $4.291\pm0.002$ \\
\hline
    $b_{112}$  & $2.8003\pm0.0005$               & $2.7871\pm0.0009$             & $2.6705\pm0.0011$             & $2.524\pm0.04$ \\
\hline
    $b_{022}$  & $7.6032\pm0.0006$               & $7.6124\pm0.0009$             & $9.1181\pm0.0012$             & $8.599\pm0.004$ \\
\hline
    $b_{103}$  & $-0.0023\pm0.0002$              & $-0.0012\pm0.0004$            & $-0.0026\pm0.0005$            & $-0.0120\pm0.0018$ \\
\hline
    $b_{013}$  &  $-0.00009\pm0.00035$           & $0.0002\pm0.006$              & $-0.0021\pm0.0007$            & $0.017\pm0.003$ \\
\hline
\end{tabular}
\caption{Coefficients of the fits for  the four definitions of asymmetry.} \label{tab:fits1}
\end{table*} 

\end{appendix}

\vskip 0.5 cm
\newpage


\end{document}